\documentclass[preprintnumbers,superscriptaddress,nofootinbib,amsmath,amssymb,aps,prd,preprint,longbibliography]{revtex4-1}
\usepackage{graphicx}
\usepackage{amsfonts}
\usepackage[colorlinks,linkcolor=red,anchorcolor=blue,citecolor=green]{hyperref}
\usepackage{subfigure}
\usepackage{natbib}
\usepackage{enumerate}
\newcommand\diff{\mathrm{d}}
\newcommand\Diff{\mathcal{D}}

\newcommand\e{\mathrm{e}}
\begin{document}
\title{
    Spherically symmetric configuration in $f(Q)$ gravity
}
\author{Rui-Hui~Lin}
\email[]{linrh@shnu.edu.cn}
\author{Xiang-Hua~Zhai}
\email[]{zhaixh@shnu.edu.cn}
\affiliation{Division of Mathematics and Theoretical Physics, Shanghai Normal University, 100 Guilin Road, Shanghai 200234, China}

\begin{abstract}
    General relativity can be formulated equivalently with a non-Riemannian geometry
    that associates with an affine connection of nonzero nonmetricity $Q$ but vanishing curvature $R$ and torsion $T$.
    Modification based on this description of gravity generates the $f(Q)$ gravity.
    In this work we explore the application of $f(Q)$ gravity to the spherically symmetric configurations.
    We discuss the gauge fixing and connections in this setting.
    We demonstrate the effects of $f(Q)$ by considering the external and internal solutions of compact stars.
    The external background solutions for any regular form of $f(Q)$ coincide with the corresponding solutions in general relativity,
    i.e., the Schwarzschild-de Sitter solution and the Reissner-Nordstr\"om-de Sitter solution with an electromagnetic field.
    For internal structure, with a simple model $f(Q)=Q+\alpha Q^2$ and a polytropic equation of state,
    we find that a negative modification ($\alpha<0$) provides support to more stellar masses
    while a positive one ($\alpha>0$) reduces the amount of matter of the star.
\end{abstract}
\maketitle

\section{Introduction}
\label{intro}
The standard description of general relativity (GR) utilizes the Riemannian geometry
and specifies the affine connection on the spacetime manifold to be metric compatible, namely the Levi-Civita one.
Yet there can be various choices of affine connection on any manifold,
and different connections may constitute different but equivalent descriptions of gravity
\cite{BeltranJimenez:2019tjy,PhysRevD.101.024053},
which may then provide different aspects of insight.
The Levi-Civita connection chosen by GR imposes that except curvature $R$,
the other two fundamental geometrical objects, the nonmetricity $Q$ and torsion $T$, should both vanish.
By loosening these conditions, one can in principle formulate theories of gravity based on
such non-Riemannian geometry that the curvature, torsion, and nonmetricity may all be nonvanishing.
Specifically, when choosing a connection requiring both curvature and nonmetricity vanish
but relaxing the constraint on torsion,
one may formulate the teleparallel equivalent of GR (TEGR) \cite{Aldrovandi:2013wha,Maluf2013}.
And as a third alternative, if one considers a flat spacetime manifold without torsion but with a nonvanishing nonmetricity,
the symmetric teleparallel formulation of GR (STGR)
\cite{Nester:1998mp,Adak:2005cd,Adak:2008gd,Mol:2014ooa,PhysRevD.97.124025,PhysRevD.98.044048,BeltranJimenez:2018vdo,PhysRevD.101.064024} may arise.

Although GR remains currently the most successful theory of gravity,
the long-known challenges of it,
including the dark contents of the Universe
and the attempts of quantizing the gravitational interaction,
still lack consensus.
The equivalent descriptions of gravity mentioned above
also inherit these issues,
which motivates the substantial study of alternative theories of gravity.
And the different geometrical bases of these theories provide novel starting points for modifications.
One of the most renowned schemes to address the dark content issue of the Universe is the
$f(R)$ gravity where the curvature scalar $R$ in the gravitational Lagrangian is replaced by some function $f(R)$
(see, e.g., Refs. \cite{DeFelice:2010aj,Sotiriou:2008rp,Capozziello:2011et,Nojiri:2010wj} for extensive reviews).
Following the same idea to consider the dark contents of the Universe as the geometric effects of the spacetime manifold,
one may start from TEGR and construct the $f(T)$ gravity
(see, e.g., Refs. \cite{PhysRevD.79.124019,PhysRevD.81.127301,Cai:2015emx,Nojiri:2017ncd}).

Similarly, by considering gravitational Lagrangian that includes an arbitrary function of the nonmetricity $Q$,
one may formulate the $f(Q)$ gravity \cite{BeltranJimenez:2018vdo}.
Like the $f(T)$ schemes, $f(Q)$ gravity also features in second order field equations instead of the fourth order ones in $f(R)$.
As one of the main motivations of this extension,
the expansion history of the Universe in $f(Q)$ gravity receives immediate attentions.
Dynamical analysis \cite{Lu:2019hra} as well as
background observation fitting by transferring to $f(z)$ \cite{PhysRevD.100.104027}
and considering cosmography \cite{PhysRevD.102.124029} has been performed.
Since at the background level, models of $f(Q)$ are indistinguishable from $f(T)$ models \cite{PhysRevD.101.103507},
Noether symmetries \cite{Dialektopoulos:2019mtr}, energy conditions \cite{PhysRevD.102.024057}, and perturbations \cite{PhysRevD.101.103507,Barros:2020bgg} in cosmological scenario are considered.
Moreover, $f(Q)$ gravity also inspires the related bouncing Universe model \cite{Bajardi:2020fxh} and several modified gravities
\cite{PhysRevD.97.124025,PhysRevD.98.084043,Xu:2019sbp,PhysRevD.101.064024,Xu:2020yeg,DAmbrosio:2020nev}.
It is safe to say that $f(Q)$ gravity is successful in explaining the accelerated expansion of the Universe
at least to the same level of statistic precision of most renowned modified gravities.
It is then natural to consider modified gravities at the astrophysical scale and attempt to break this so-called degeneracy.
The growing detections of gravitational waves \cite{PhysRevLett.116.061102}
and the direct observation of the center of our Galaxy \cite{Akiyama:2019cqa}
seem to be promising tools in this regard.

In the context of $f(R)$ gravity, spherically symmetric solutions including black holes
(see, e.g., primarily Refs. \cite{PhysRevD.74.064022,PhysRevD.80.124011})
and stars (see, e.g., primarily Refs. \cite{PhysRevD.78.064019,PhysRevD.80.064002}) are extensively studied.
The same symmetry is also investigated under the $f(T)$ framework
(see, e.g., Refs. \cite{PhysRevD.90.124006,PhysRevD.94.124025,Lin:2016xyi,Krssak:2015oua,PhysRevD.98.064047,Bahamonde:2019jkf}).
In this paper, we intend to study the spherically symmetric configurations in $f(Q)$ gravity.
We discuss the gauge fixing and obtain the components of connection for this scenario.
We then explore the external and internal solutions of compact stars in $f(Q)$ gravity.
The paper is organized as follows.
In Sec \ref{fQreview}, we briefly review the nonmetricity description of GR and the $f(Q)$ gravity.
The gauge-fixing issue and the covariant formulation of $f(Q)$ gravity are investigated in Sec \ref{gaugefixing}.
Section \ref{sph} focuses on the spherical settings of $f(Q)$ gravity.
We discuss the solutions of external spacetime and interior structure of compact objects in Secs \ref{external} and \ref{stars}, respectively.
Section \ref{conclusion} contains our concise summary and discussions.
Throughout the paper we will be working in the natural unit where the speed of light $c=1$.

\section{Nonmetricity and $f(Q)$ gravity}
\label{fQreview}
As the spacetime manifold $\mathcal{M}$ is assumed to be a parallelizable and differentiable metric space,
the geometry of it is generally determined by a Lorentzian metric $g$ and a linear connection $\Gamma$.
The metric $g$ is conveniently written in terms of coframe 1-forms
\begin{equation}
    \label{metric}
    g=g_{\alpha\beta}\diff x^\alpha\otimes\diff x^\beta=\eta_{ab}\vartheta^a\otimes\vartheta^b,
\end{equation}
where $\eta$ is the Minkowski metric,
and $\{\vartheta^a\}$ is the dual coframe of a general frame $\{e_a\}$ with the relation
\begin{equation}
    \vartheta^b(e_a)=\delta^b_a.
\end{equation}
The full connection 1-form $\Gamma^a_{\;\:b}$ is related to a covariant derivative $\Diff(\Gamma)$
and can be decomposed uniquely as \cite{Ortin:2004ms}
\begin{equation}
    \label{connection}
    \Gamma^a_{\;\:b}=\omega^a_{\;\:b}+K^a_{\;\:b}+L^a_{\;\:b},
\end{equation}
where $\omega^a_{\;\:b}$ is the Levi-Civita connection 1-form with
\begin{equation}
    \label{covderw}
    \Diff(\omega) \vartheta^a=d \vartheta^a+\omega^a_{\;\:b}\wedge \vartheta^b=0,
\end{equation}
$K^a_{\;\:b}$ is the contorsion 1-form,
and $L^a_{\;\:b}$ is the deformation 1-form.
Here $d$ is the exterior derivative.
In terms of components, Eq. \eqref{connection} is written as
\begin{equation}
    \label{connectioncom}
    \Gamma^\alpha_{\beta\gamma}=\{^\alpha_{\beta\gamma}\}+K^\alpha_{\beta\gamma}+L^\alpha_{\beta\gamma}.
\end{equation}
The nonmetricity 1-form given by this connection is
\begin{equation}
    \label{nonmetric0}
    \begin{split}
        Q_{ab}=&\frac12\Diff(\Gamma)\eta_{ab}=\Gamma_{(ab)}=-A^c_{\:\;b}\eta_{ac}-A^c_{\:\;a}\eta_{cb},
    \end{split}
\end{equation}
where $A^a_{\;\:b}=K^a_{\;\:b}+L^a_{\;\:b}$ and $\Diff(\omega)\eta_{ab}=0$ is utilized.
If $K^a_{\;\:b}$ vanishes, then the nonmetricity can be written in terms of components
\begin{equation}
    \label{Qcom}
    Q_{\alpha\beta\gamma}=\nabla_\alpha g_{\beta\gamma}=-L^\rho_{\alpha\beta}g_{\rho\gamma}-L^\rho_{\alpha\gamma}g_{\rho\beta},
\end{equation}
or equivalently,
\begin{equation}
    \label{Qcom1}
    L^\alpha_{\beta\gamma}=\frac12Q^\alpha_{\beta\gamma}-Q_{(\beta\gamma)}^{\quad\;\alpha},
\end{equation}
where $\nabla_\alpha$ denotes the components of $\mathcal D(\Gamma)$.
The torsion 2-form and curvature 2-form are given by
\begin{equation}
    \label{torsion0}
    T^a=\Diff(\Gamma)\vartheta^a=d\vartheta^a+\Gamma^a_{\;\:b}\wedge \vartheta^b=A^a_{\;\:b}\wedge \vartheta^b,
\end{equation}
and
\begin{equation}
    \label{curvature0}
    R^a_{\;\:b}(\Gamma)=\Diff(\Gamma)\Gamma^a_{\;\:b}=d\Gamma^a_{\;\:b}+\Gamma^a_{\;\:c}\wedge\Gamma^c_{\;\:b},
\end{equation}
respectively.
If $\Gamma^a_{\;\:b}=\omega^a_{\;\:b}$, then the Riemannian curvature is given by
\begin{equation}
    \label{curvature1}
    R^a_{\;\:b}(\omega)=d\omega^a_{\;\:b}+\omega^a_{\;\:c}\wedge\omega^c_{\;\:b}.
\end{equation}
Thus,
\begin{equation}
    \label{curvature2}
    R^a_{\;\:b}(\Gamma)=R^a_{\;\:b}(\omega)+\Diff(\omega)A^a_{\;\:b}+A^a_{\;\:c}\wedge A^c_{\;\:b}.
\end{equation}
Let $h^{ab\cdots}=*(\vartheta^a\wedge\vartheta^b\wedge\cdots)$, then the curvature 4-form is
\begin{equation}
    \label{curvaturescalar0}
    \begin{split}
        R(\Gamma)=&R^a_{\;\:b}\wedge h^b_{\;\:a}\\
        =&R(\omega)+\Diff(\omega)A^a_{\;\:b}\wedge h^b_{\;\:a}+A^a_{\;\:c}\wedge A^c_{\;\:b}\wedge h^b_{\;\:a}\\
        =&R(\omega)+d(A^a_{\;\:b}\wedge h^b_{\;\:a})-A^a_{\;\:b}\wedge\Diff(\omega)h^b_{\;\:a}+A^a_{\;\:c}\wedge A^c_{\;\:b}\wedge h^b_{\;\:a}.
    \end{split}
\end{equation}
Note that $R(\omega)$ constitutes the Hilbert-Einstein Lagrangian 4-form of the standard formulation of GR.
If one wishes to describe the gravitation with only the nonmetricity and imposes on the connection that
it should be torsionless and curvatureless, i.e., both $T^a(\Gamma)$ and $R^a_{\;\:b}(\Gamma)$ vanish
\cite{Nester:1998mp,Adak:2005cd,Adak:2008gd,PhysRevD.98.044048}, then
\begin{equation}
    \label{HELag}
    \begin{split}
        \mathcal L_\text{EH}=&\frac1{2\kappa}R(\omega)\\
        =&-d(\frac1{2\kappa}L^a_{\;\:b}\wedge h^b_{\;\:a})+\frac1{2\kappa}L^a_{\;\:c}\wedge L^c_{\;\:b}\wedge h^b_{\;\:a},
    \end{split}
\end{equation}
where $\kappa=8\pi G$ and $G$ is the Newtonian gravitation constant,
and $\Diff(\omega)h^b_{\;\:a}=0$ has been utilized.
Discarding the exact form and defining the nonmetricity 4-form $\mathcal Q=L^a_{\;\:c}\wedge L^c_{\;\:b}\wedge h^b_{\;\:a}$,
one thus arrives at the Lagrangian 4-form of STGR
\begin{equation}
    \label{STGRLag}
    \mathcal L_\text{STGR}=\frac1{2\kappa}L^a_{\;\:c}\wedge L^c_{\;\:b}\wedge h^b_{\;\:a}=\frac1{2\kappa}\mathcal Q.
\end{equation}
In terms of component, it is written as
\begin{equation}
    \label{STGRLag1}
    \mathcal L_\text{STGR}=-\frac{\sqrt{-g}}{2\kappa}g^{\mu\nu}\left( L^\alpha_{\beta\nu}L^\beta_{\mu\alpha}-L^\beta_{\alpha\beta}L^\alpha_{\mu\nu} \right)\diff^4 x=\frac{\sqrt{-g}}{2\kappa}Q\diff^4x,
\end{equation}
where the nonmetricity scalar $Q\equiv -g^{\mu\nu}\left( L^\alpha_{\beta\nu}L^\beta_{\mu\alpha}-L^\beta_{\alpha\beta}L^\alpha_{\mu\nu} \right)$.
Utilizing the nonmetricity conjugate
\begin{equation}
    \label{Pdef}
    P^\alpha_{\;\beta\gamma}=-\frac12L^\alpha_{\beta\gamma}+\frac{1}{4}\left( Q^\alpha-\tilde{Q}^\alpha \right)g_{\beta\gamma}-\frac14\delta^\alpha_{(\beta}Q_{\gamma)},
\end{equation}
where $Q_\alpha=Q_{\alpha\beta}^{\quad\beta}$ and $\tilde{Q}_\alpha=Q^\beta_{\;\alpha\beta}$
are the two independent traces of the nonmetricity tensor,
one finds that
\begin{equation}
    \label{Qdef1}
    \begin{split}
        Q=&-g^{\mu\nu}\left( L^\alpha_{\beta\nu}L^\beta_{\mu\alpha}-L^\beta_{\alpha\beta}L^\alpha_{\mu\nu} \right)\\
        =&-P^{\alpha\beta\gamma}Q_{\alpha\beta\gamma}.
    \end{split}
\end{equation}

Variation of Eq. \eqref{STGRLag} with respect to metric $g$ gives \cite{Adak:2005cd,Mol:2014ooa}
\begin{equation}
    \label{eomSTGRd}
    \left[ \Diff(\omega)\left( L^{ab}-Q^{ab} \right)+L^a_{\;c}\wedge L^{cb} \right]\wedge h_{kab}=2\kappa\tau_k,
\end{equation}
where $\tau_k$ is the energy-momentum 3-form.
In terms of components, Eq. \eqref{eomSTGRd} can be written as
\begin{equation}
    \label{eomSTGR}
    \begin{split}
        &2\bar{\nabla}_\alpha \left( P^\alpha_{\;\mu\nu} \right)+\left( L^\alpha_{\beta\nu}L^\beta_{\mu\alpha}-L^\beta_{\alpha\beta}L^\alpha_{\mu\nu} \right)+\frac12g_{\mu\nu}Q=\kappa\mathcal{T}_{\mu\nu},
    \end{split}
\end{equation}
where $\bar\nabla_\alpha$ denotes the component of $\Diff(\omega)$,
and $\mathcal{T}_{\mu\nu}$ is the energy-momentum tensor.

In $f(Q)$, gravity one usually replaces $Q$ in the Lagrangian \eqref{STGRLag1} with an arbitrary function $f(Q)$ and considers the gravitational action
\begin{equation}
    \label{action}
    \mathcal{S}=\frac1{2\kappa}\int\sqrt{-g}f(Q)\diff^4x.
\end{equation}
Then the field equation of $f(Q)$ gravity becomes
\begin{equation}
    \label{eomfQ}
    \begin{split}
        &2\bar{\nabla}_\alpha \left( f_QP^\alpha_{\;\mu\nu} \right)+f_Q \left( L^\alpha_{\beta\nu}L^\beta_{\mu\alpha}-L^\beta_{\alpha\beta}L^\alpha_{\mu\nu} \right)+\frac12g_{\mu\nu}f=\kappa\mathcal{T}_{\mu\nu}.
    \end{split}
\end{equation}
Here and thereafter, we denote $f_Q=\diff f(Q)/\diff Q$ and $f_{QQ}=\diff^2 f(Q)/\diff Q^2$.
As a more unifying form using $\nabla_\alpha$ instead of $\bar\nabla_\alpha$,
Eq. \eqref{eomfQ} can be written as
\begin{equation}
    \label{eomfQ4}
    -\frac2{\sqrt{-g}}\nabla_\alpha \left( \sqrt{-g}f_QP^\alpha_{\;\mu\nu} \right)+f_Q \left( P_\nu^{\;\alpha\beta}Q_{\mu\alpha\beta}-2P^{\alpha\beta}_{\;\:\;\:\mu}Q_{\alpha\beta\nu} \right)+\frac12g_{\mu\nu}f=\kappa\mathcal{T}_{\mu\nu}.
\end{equation}

From Eq. \eqref{curvature2}, the vanishing of $R^a_{\;\:b}(\Gamma)$ leads to
\begin{equation}
    \label{curvature4}
    R^a_{\;\:b}(\omega)=-\Diff(\omega)A^a_{\;\:b}-A^a_{\;\:c}\wedge A^c_{\;\:b}.
\end{equation}
With $P^{\alpha\rho}_{\quad\rho}=\frac12 \left( Q^\alpha-\tilde{Q}^\alpha \right)$,
the components of Riemannian curvature, Ricci tensor and Ricci scalar are
\begin{equation}
    \label{curvature5}
    \begin{split}
        \bar R^\rho_{\;\:\mu\lambda\nu}=&\bar\nabla_\nu L^\rho_{\mu\lambda}-\bar\nabla_\lambda L^\rho_{\mu\nu}+L^\rho_{\sigma\nu}L^\sigma_{\mu\lambda}-L^\rho_{\sigma\lambda}L^\sigma_{\mu\nu},\\
        \bar R_{\mu\nu}=&\bar\nabla_\nu L^\rho_{\mu\rho}-\bar\nabla_\rho L^\rho_{\mu\nu}+L^\rho_{\sigma\nu}L^\sigma_{\mu\rho}-L^\rho_{\sigma\rho}L^\sigma_{\mu\nu}\\
        =&2\bar\nabla_\rho P^\rho_{\;\:\mu\nu}-\frac12\bar\nabla_\rho \left( Q^\rho-\tilde{Q}^\rho \right)g_{\mu\nu}+L^\rho_{\sigma\nu}L^\sigma_{\mu\rho}-L^\rho_{\sigma\rho}L^\sigma_{\mu\nu},\\
        \bar R=&-\bar\nabla_\rho \left( Q^\rho-\tilde{Q}^\rho \right)-Q,
    \end{split}
\end{equation}
respectively.
Then Eq. \eqref{eomfQ} can also be written as
\begin{equation}
    \label{eomfQ3}
    2f_{QQ}P^\alpha_{\;\:\mu\nu}\partial_\alpha Q+\frac12 g_{\mu\nu}\left( f-f_QQ \right)+f_Q G_{\mu\nu}=\kappa\mathcal T_{\mu\nu}.
\end{equation}
One can easily see that Einstein equation is recovered when $f(Q)=Q$.

\section{The covariant formulation of $f(Q)$ gravity}
\label{gaugefixing}
As mentioned in the Introduction, the connection of the geometry is considered to be curvatureless and torsionless under the current framework.
Generally, the full connection $\Gamma$ cannot be solved uniquely from $R^a_{\;\:b}(\Gamma)=0$ and $T^a(\Gamma)=0$.
Thus, it is necessary to develop a prescription to determine $\Gamma$ or its components in any given frame
before proceeding to consider the spherically symmetric cases of the current work.

As discussed in Ref. \cite{Adak:2008gd}, it is always possible to choose $\Gamma^a_{\;\:b}=0$,
such that, together with the vanishing of $K^a_{\;\:b}$, both $R^a_{\;\:b}(\Gamma)=0$ and $T^a(\Gamma)=0$ can be satisfied.
The same gauge choice is also used in various studies of STGR \cite{Nester:1998mp,Adak:2005cd,Mol:2014ooa}
and is dubbed as the coincident gauge in further investigations of STGR \cite{PhysRevD.98.044048,BeltranJimenez:2019tjy}.
Note that $K^a_{\;\:b}$ is a tensor, so its components remain zero in any frame.
$\Gamma^a_{\;\:b}$, on the other hand, is not a tensor,
and hence $\Gamma^a_{\;\:b}=0$ does not mean its components $\Gamma^\alpha_{\beta\gamma}$ in Eq. \eqref{connectioncom} will always vanish.
Therefore, $\Gamma^\alpha_{\beta\gamma}=0$, and hence $L^\alpha_{\beta\gamma}=-\left\{ ^\alpha_{\beta\gamma} \right\}$,
can only be satisfied in a certain class of frames.
For any diffeomorphism or frame transformation $\chi(x)$,
the components of $\Gamma$ can be determined by \cite{PhysRevD.98.044048}
\begin{equation}
    \label{coordtransf}
    \Gamma^\alpha_{\beta\gamma}=\frac{\partial x^\alpha}{\partial \chi^\rho}\partial_\beta\partial_\gamma\chi^\rho.
\end{equation}
This transformation of $\Gamma^\alpha_{\beta\gamma}$ thus can be viewed as the core of the covariant formulation of $f(Q)$ gravity,
which preserves the Lorentz covariance under the nonmetricity framework.
It then comes to the crucial problem of determining the frame in which one should set $\Gamma^\alpha_{\beta\gamma}=0$.
The $f(T)$ counterpart provides valuable insights \cite{Krssak:2015oua}.
For local Lorentz transformation $\chi^\mu=\Lambda^\mu_\nu x^\nu$ Eq. \eqref{coordtransf} reads
\begin{equation}
    \label{lorentztransf}
    \Gamma^\alpha_{\beta\gamma}=\Lambda^\alpha_\rho\partial_\beta \left( \Lambda^{-1} \right)^\rho_\gamma,
\end{equation}
which indicates that the total connection $\Gamma$ is purely inertial.
Indeed, under the nonmetricity framework,
the tensor $Q_{\alpha\beta\gamma}$ is considered to encode all and only the information of gravity.
It follows that all components of $Q_{\alpha\beta\gamma}$ should vanish when gravity is \textit{canceled},
denoted as $Q_{\alpha\beta\gamma}|_{G=0}=0$.
Then,
\begin{equation}
    \label{Lcancel}
    \left. L^\alpha_{\beta\gamma}\right|_{G=0}=\left.\left( \frac12Q^\alpha_{\beta\gamma}-Q_{(\beta\gamma)}^{\quad\;\alpha} \right)\right|_{G=0}=0,
\end{equation}
and hence
\begin{equation}
    \label{Gammacancel}
    \left. \Gamma^\alpha_{\beta\gamma}\right|_{G=0}=\left.\left\{ ^\alpha_{\beta\gamma} \right\}\right|_{G=0}+\left. L^\alpha_{\beta\gamma}\right|_{G=0}=\left.\left\{ ^\alpha_{\beta\gamma} \right\}\right|_{G=0}.
\end{equation}
Furthermore, since $R^a_{\;\:b}(\Gamma)=0$, $T^a(\Gamma)=0$ and gravity is all encoded in $Q_{\alpha\beta\gamma}$,
the purely inertial connection $\Gamma^\alpha_{\beta\gamma}$ should remain the same whether the gravity is canceled or not, i.e.,
\begin{equation}
    \label{Gammacancel1}
    \Gamma^\alpha_{\beta\gamma}=\left. \Gamma^\alpha_{\beta\gamma}\right|_{G=0}=\left.\left\{ ^\alpha_{\beta\gamma} \right\}\right|_{G=0}.
\end{equation}
This gives a prescription to determine the components of $\Gamma$.
As a first example, the Cartesian frame gives
$\Gamma^\alpha_{\beta\gamma}=\left.\left\{ ^\alpha_{\beta\gamma} \right\}\right|_{G=0}=0$,
which is usually seen in the spatial-flat case of cosmology.
For any other frame,
the components of the connection $\Gamma^\alpha_{\beta\gamma}$
can be obtained either via Eq. \eqref{coordtransf} or \eqref{lorentztransf} from Cartesian frame,
or via direct calculation of $\left.\left\{ ^\alpha_{\beta\gamma} \right\}\right|_{G=0}$.

\section{$f(Q)$ gravity with spherical symmetry}
\label{sph}
We now proceed to consider the spherically symmetric configurations.
The metric takes the usual form in the spherical coordinates
\begin{equation}
    \label{ssmetric}
    \diff s^2=\e^\xi\diff t^2-\e^\zeta\diff r^2-r^2 \left( \diff\theta^2+\sin^2\theta\diff\phi^2 \right),
\end{equation}
where $\xi=\xi(r)$ and $\zeta=\zeta(r)$ are functions depending only on the radial coordinate $r$.
In this coordinate system, the nonvanishing components of $\Gamma$ are given by Eq. \eqref{Gammacancel1},
\begin{equation}
    \label{gamma}
    \begin{split}
        &\Gamma^r_{\phi\phi}=\sin^2\theta\Gamma^r_{\theta\theta}=-r\sin^2\theta,\\
        &\Gamma^\theta_{r\theta}=\Gamma^\phi_{r\phi}=\frac1r,\\
        &\Gamma^\theta_{\phi\phi}=-\cos\theta\sin\theta,\quad\Gamma^\phi_{\theta\phi}=\cot\theta.
    \end{split}
\end{equation}
Therefore, the nonvanishing $Q_{\alpha\beta\gamma}$ and $L^\alpha_{\beta\gamma}$ are
\begin{equation}
    \label{Qcomponents}
    \begin{split}
        &Q_{rtt}=\e^\xi\xi',\quad Q_{rrr}=-\e^\zeta\zeta',\\
        &Q_{\theta r\theta}=Q_{\theta\theta r}=r-\e^\zeta r,\\
        &Q_{\phi r\phi}=Q_{\phi\phi r}=\left( 1-\e^\zeta \right)r\sin^2\theta,
    \end{split}
\end{equation}
and
\begin{equation}
    \label{Lcomponents}
    \begin{split}
        &L^t_{tr}=L^t_{rt}=-\frac12\xi',\quad L^r_{rr}=-\frac12\zeta',\\
        &L^r_{tt}=-\frac12\e^{\xi-\zeta}\xi',\quad L^r_{\theta\theta}=\left( \e^{-\zeta}-1 \right)r,\\
        &L^r_{\phi\phi}=\e^{-\zeta}\left( 1-\e^{\zeta} \right)r\sin^2\theta,
    \end{split}
\end{equation}
respectively, where $'$ indicates derivative with respect to $r$.
And the nonmetricity scalar $Q$ is
\begin{equation}
    \label{Qsc}
    Q=\frac{\left( \e^{-\zeta}-1 \right)\left( \zeta'+\xi' \right)}{r}.
\end{equation}
The components of Eq. \eqref{eomfQ} read
\begin{equation}
    \label{eomsph}
    \begin{split}
        \kappa\mathcal{T}_{tt}=&\frac{\e^{\xi-\zeta}}{2r^2}\left\{ 2rf_{QQ}Q'\left( \e^\zeta -1 \right)+f_Q \left[ \left( \e^\zeta-1 \right)\left( 2+r\xi' \right)+\left( 1+\e^\zeta \right)r\zeta' \right]+fr^2\e^\zeta \right\},\\
        \kappa\mathcal{T}_{rr}=&-\frac{1}{2r^2}\left\{ 2rf_{QQ}Q'\left( \e^\zeta -1 \right)+f_Q \left[ \left( \e^\zeta-1 \right)\left( 2+r\zeta'+r\xi' \right)-2r\xi' \right]+fr^2\e^\zeta \right\},\\
        \kappa\mathcal{T}_{\theta\theta}=&-\frac{r}{4\e^\zeta}\left\{ -2rf_{QQ}Q'+f_Q \left[ 2\xi'\left( \e^\zeta-2 \right)-r\xi'^2+\zeta'\left( 2\e^\zeta+r\xi' \right)-2r\xi'' \right]+fr^2\e^\zeta \right\}.
    \end{split}
\end{equation}

\section{External solutions}
\label{external}
We first study the spherical vacuum with $\mathcal{T}_{\mu\nu}=0$
and search for spherical black hole solutions of $f(Q)$ gravity.
Consider a linear combination of the $tt$ and $rr$ components of Eq. \eqref{eomsph},
\begin{equation}
    \label{combinevac}
    0=\e^\zeta\mathcal{T}_{tt}+\e^\xi\mathcal{T}_{rr}=\frac{\e^\xi f_Q \left( \zeta'+\xi' \right)}{r}.
\end{equation}
It follows that $\xi'+\zeta'=0$, which determines $g_{rr}=g_{tt}^{\;\;-1}$ up to a constant rescaling of the temporal coordinate.
Comparing with Eq. \eqref{Qsc}, one finds that for a spherical vacuum,
the nonmetricity scalar $Q$ always vanishes.
Equation \eqref{eomsph} then becomes
\begin{equation}
    \label{eomvac1}
    2f_{Q0}\left( \e^\zeta-1+r\zeta' \right)+f_0r^2\e^\zeta=0,
\end{equation}
where $\left.f_0=f(Q)\right|_{Q=0}$ and $\left.f_{Q0}=f_Q\right|_{Q=0}$ are constants.
It can be solved directly as
\begin{equation}
    \label{solvac}
    \e^\xi=\e^{-\zeta}=1+\frac{r_\text s}r+\frac{f_0r^2}{6f_{Q0}},
\end{equation}
where $r_\text s$ is an integral constant.
This is exactly the Schwarzschild-de Sitter (SdS) solution with the cosmological constant $\Lambda=f_0/(2f_{Q0})$.
This suggests that for any form of $f(Q)$, as long as $f(Q)$ and $f_Q$ are regular at $Q=0$,
the static spherical vacuum solutions can only be the same ones as in GR.
This can also be seen from Eq. \eqref{eomfQ3}, which becomes
\begin{equation}
    \label{eomfQ2}
    G_{\mu\nu}+\frac{f_0}{2f_{Q0}}g_{\mu\nu}=0,
\end{equation}
when $Q=0$.

For another configuration, we consider the external solution with the energy-momentum tensor of an electromagnetic field,
\begin{equation}
    \label{emem}
    \mathcal T_{\mu\nu}=F_\mu^{\;\:\alpha}F_{\nu\alpha}-\frac14g_{\mu\nu}F_{\alpha\beta}F^{\alpha\beta},
\end{equation}
where $F_{\alpha\beta}$ is the field strength tensor.
The solution in this case may represent the external of a charged spherical black hole.
The Maxwell's equation $dF=0$ still gives $F_{rt}=\frac C{r^2}\sqrt{\e^{\xi+\zeta}}$
as in the standard GR, where $C$ is an integral constant, and
\begin{equation}
    \label{emem2}
    \mathcal T_{tt}=\frac{C^2}{2r^4}\e^\xi,\quad\mathcal T_{rr}=-\frac{C^2}{2r^4}\e^\zeta.
\end{equation}
It follows that Eq. \eqref{combinevac} still holds,
and hence $Q=0$ and $\xi$ can be chosen as $\xi=-\zeta$.
Equation \eqref{eomsph} then becomes
\begin{equation}
    \label{eomem1}
    2f_{Q0}\left( \e^\zeta-1+r\zeta' \right)+f_0r^2\e^\zeta=\frac{\kappa C^2}{r^2}\e^\zeta,
\end{equation}
which can be solved directly as
\begin{equation}
    \label{solem}
    \e^\xi=\e^{-\zeta}=1+\frac{r_\text s}r+\frac{\kappa C^2}{2f_{Q0}r^2}+\frac{f_0r^2}{6f_{Q0}}.
\end{equation}
Again, this is exactly the Reissner-Nordstr\"om-de Sitter (RNdS) solution.
And when $Q=0$, Eq. \eqref{eomfQ3} becomes
\begin{equation}
    \label{eomfQ5}
    G_{\mu\nu}+\frac{f_0}{2f_{Q0}}g_{\mu\nu}=\frac{\kappa}{2f_{Q0} }\mathcal T_{\mu\nu},
\end{equation}
which can be viewed as the Einstein equation with a cosmological constant $\Lambda=f_0/(2f_{Q0})$ and an electromagnetic field.

For comparison, it is worth noting that in $f(R)$ gravity,
although SdS and RNdS solutions are still viable when $R=\text{const.}$,
external solutions generally depend on the concrete forms of $f(R)$ (see, e.g., Ref. \cite{Capozziello:2011et}).
In $f(T)$ gravity, depending on the choice of proper tetrad,
the external solutions may vary with the different forms of the function $f(T)$ \cite{PhysRevD.94.124025,Bahamonde:2019jkf}
or remain the same as in GR due to an always constant torsion scalar $T$ \cite{Lin:2019tos}.
It seems that despite the different geometrical bases,
when the corresponding geometrical object,
i.e. the curvature scalar $R$, torsion scalar $T$, or nonmetricity scalar $Q$,
takes a constant value with respect to the radial coordinate,
$f(R)$, $f(T)$, and $f(Q)$ gravities will result in the identical external solutions of GR (or TEGR/STGR).
According to the analysis in this section,
as $Q$ is always a constant function regardless of the concrete form of $f(Q)$,
it is natural that only the SdS and RNdS solutions remain.

Nonetheless, within the framework of nonmetricity,
conceptually it can be argued that the so-called cosmological constant is given by the modification from the function $f(Q)$,
and the charge is also altered by it.
Moreover, dynamically the external solutions in any $f(Q)$ gravity are not necessarily trivial.
Consider a perturbation of the energy-momentum tensor
\begin{equation}
    \label{pert}
    \tilde{\mathcal{T}}_{\mu\nu}=\mathcal{T}_{\mu\nu}+\epsilon\mathcal{T}^{(1)}_{\mu\nu}
\end{equation}
with an infinitesimal $\epsilon$,
which leads to the perturbations of the metric and the Einstein tensor
\begin{equation}
    \label{pertgG}
    \tilde{g}_{\mu\nu}=g_{\mu\nu}+\epsilon g^{(1)}_{\mu\nu}+\mathcal O(\epsilon^2),\quad\tilde{G}_{\mu\nu}=G_{\mu\nu}+\epsilon G^{(1)}_{\mu\nu}+\mathcal O(\epsilon^2),
\end{equation}
as well as the nonmetricity conjugate tensor and the nonmetricity scalar,
\begin{equation}
    \label{pertPQ}
    \tilde{P}^\alpha_{\;\:\mu\nu}=P^\alpha_{\;\:\mu\nu}+\epsilon P^{(1)\alpha}_{\;\:\;\:\mu\nu}+\mathcal O(\epsilon^2),\quad\tilde{Q}=Q+\epsilon Q^{(1)}+\mathcal O(\epsilon^2).
\end{equation}
Then the first order perturbation of Eq. \eqref{eomfQ3} near the background $Q=0$ is
\begin{equation}
    \label{perteom}
    2f_{QQ}P^\alpha_{\mu\nu}\partial_\alpha Q^{(1)}+\frac12g^{(1)}_{\mu\nu}\left( f-f_QQ \right)+f_QG^{(1)}_{\mu\nu}=\kappa\mathcal T^{(1)}_{\mu\nu}.
\end{equation}
Thus, at perturbation level, $f(Q)$ is generally different from GR depending on the specific form of $f$.
A similar situation is also seen in $f(R)$ gravity when $R=\text{const.}$ \cite{PhysRevLett.101.099001}.
In $f(T)$ gravity, if one adopts such a choice of proper tetrad that the torsion scalar $T=\text{const.}$
and that the external solutions are also identical to GR \cite{Lin:2019tos},
then the perturbative equation can be written in a similar form
\begin{equation}
    \label{perteomt}
    2f_{TT}S_{\mu\nu}^{\;\:\;\:\alpha}\partial_\alpha T^{(1)}+\frac12g^{(1)}_{\mu\nu}\left( f-f_TT \right)+f_TG^{(1)}_{\mu\nu}=\kappa\mathcal T^{(1)}_{\mu\nu}.
\end{equation}
Therefore, the situation in $f(T)$ gravity also appears to be similar.

\section{Internal solutions}
\label{stars}
In this section, we explore the interior structure of a spherically symmetric object under the framework of $f(Q)$ gravity.
For a perfect fluid in a comoving frame, the energy-momentum tensor $\mathcal T_{\mu\nu}$ can be written as
\begin{equation}
    \label{perfectfluid}
    \mathcal T_{\mu\nu}=\left( \rho+p \right)u_\mu u_\nu-pg_{\mu\nu},
\end{equation}
where $\rho$ and $p$ are the energy density and pressure, respectively,
and $u^\mu$ is the 4-velocity with $u^\mu u_\mu=1$.
For stars within which gravitational forces are supported by nuclear reaction,
the interior balance can be described by Newtonian approximation and the Lane-Emden equation.
For other more compact stars that are mostly supported by the degeneracy pressure of nuclear particles,
a relativistic treatment is required.
In the standard GR, this is described by the Tolman-Oppenheimer-Volkoff (TOV) equation.
In $f(Q)$ gravity, this is determined by Eq. \eqref{eomsph} as well as the conservation of $\mathcal T_{\mu\nu}$,
\begin{equation}
    \label{dp}
    \frac{\diff p}{\diff r}=-\frac12(p+\rho)\xi'.
\end{equation}

In order to solve Eqs. \eqref{dp} and \eqref{eomsph} for the metric and the profiles of pressure and energy density,
inputs of a specific $f(Q)$ and matter equation of state (EOS) are needed.
Moreover, the viable and accustomed external solution can provide junction conditions for this differential equation system.
Specifically, as discussed in $f(R)$ gravity \cite{Deruelle:2007pt,Senovilla:2013vra},
the metric of the spacetime manifold should not involve a problematic term with a $\delta$ function
at the boundary of the compact star so that there is no brane there.
The field equation \eqref{eomfQ3} still has a term proportional to $G_{\mu\nu}$.
Thus, the usual GR junction conditions , i.e., $\left[ h_{\mu\nu} \right]=0$ and $\left[ \mathcal K \right]=0$,
are still required, where $\left[ \; \right]$ indicates the jump across the surface,
$h_{\mu\nu}$ and $\mathcal K$ are the induced metric and the extrinsic curvature of the hyper surface at the boundary, respectively.
In addition to that, the first term of Eq. \eqref{eomfQ3} requires that $\left[ Q \right]=0$
so that $\partial_r Q$ does not lead to a $\delta$ function, either.
Therefore, a well-defined external metric is needed for the internal solution to match.
This requirement is easily met in $f(Q)$ gravity since the external solutions are identical to those of GR.

As a heuristic example, we consider a simple model
\begin{equation}
    \label{fQform}
    f(Q)=Q+\alpha Q^2,
\end{equation}
where $\alpha$ is the model parameter with the dimension of the square of length.
It can be viewed as an analog to the Starobinsky model in curvature-based gravity \cite{Starobinsky:1980te}.
For the EOS of stellar matter, we adopt the polytropic approximation
\begin{equation}
    \label{poly}
    p=k\rho^\gamma,
\end{equation}
where for interacting Fermi gas we have $k\simeq 2.0\times10^5\:\text{cm}^5\cdot\text g^{-1}\cdot\text s^{-2}$ when $\gamma=2$ \cite{prakash}.
Note that $\xi(r)$ does not appear in the equations
(after the common factor is eliminated).
We therefore solve the system for the functions $\xi'(r),\zeta(r)$ and $\rho(r)$.
The regularity conditions at the center of the star give one of the initial conditions that $\xi'(0)=0$.
The direct joining of the internal metric with the well-defined exterior SdS metric admits the Arnowitt-Deser-Misner mass,
which, for comparison, is not available in $f(T)$ gravity \cite{PhysRevD.98.064047}.
Then the vanishing of mass at the center demands that $\zeta(0)=0$.
The radius $\mathcal R$ of the star is defined by
\begin{equation}
    \label{bc2}
    \rho(\mathcal R)=0.
\end{equation}
The exact value $\xi(0)$ is determined by meeting the junction condition at $\mathcal R$.

\begin{figure}[htpb]
    \centering
    \includegraphics[width=0.8\linewidth]{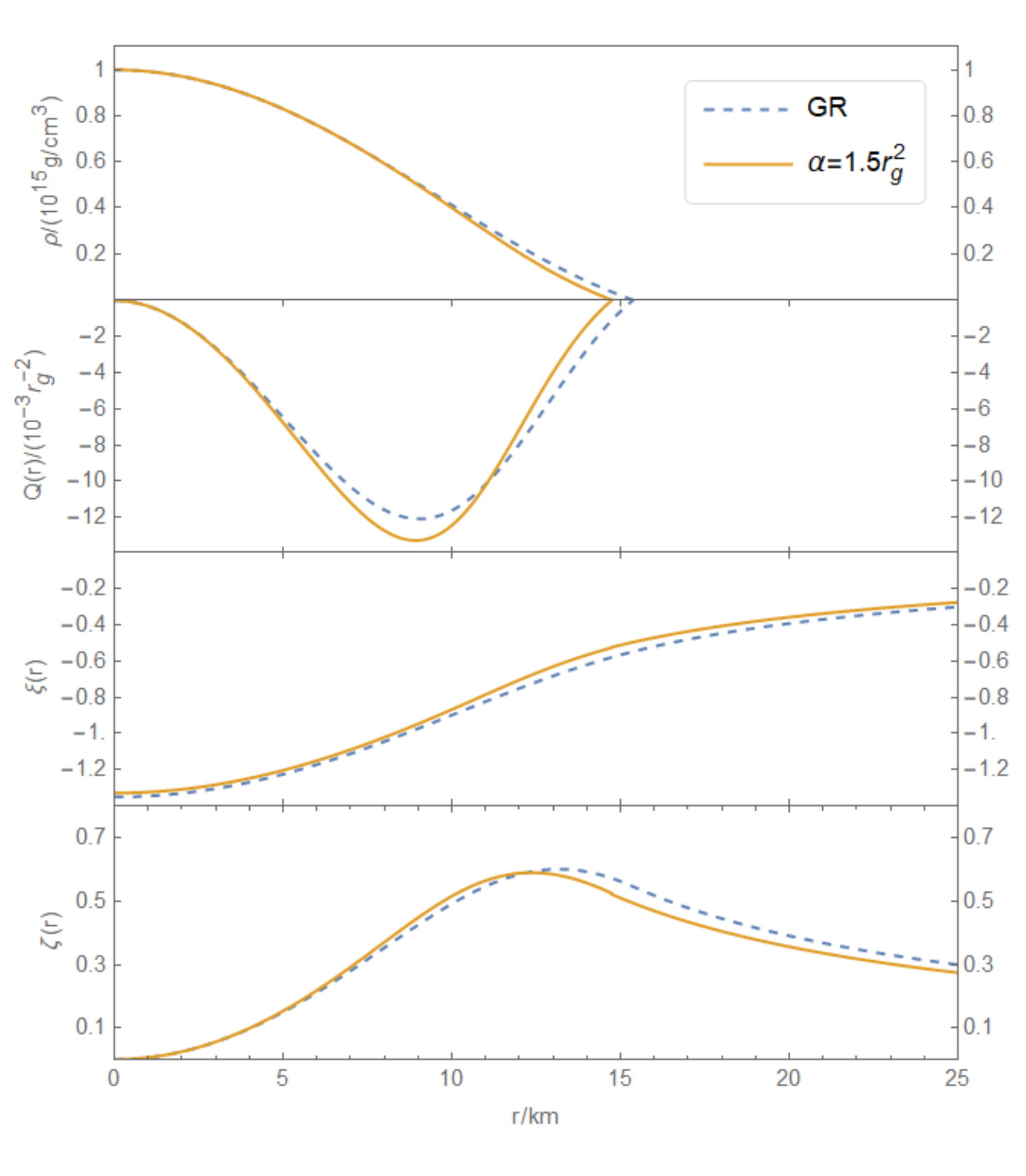}
    \caption{The profiles of the energy density $\rho$,
        nonmetricity scalar $Q$, and the metric $\xi,\zeta$.}
    \label{fig:profile}
\end{figure}

\begin{figure}[htpb]
    \centering
    \includegraphics[width=0.5\linewidth]{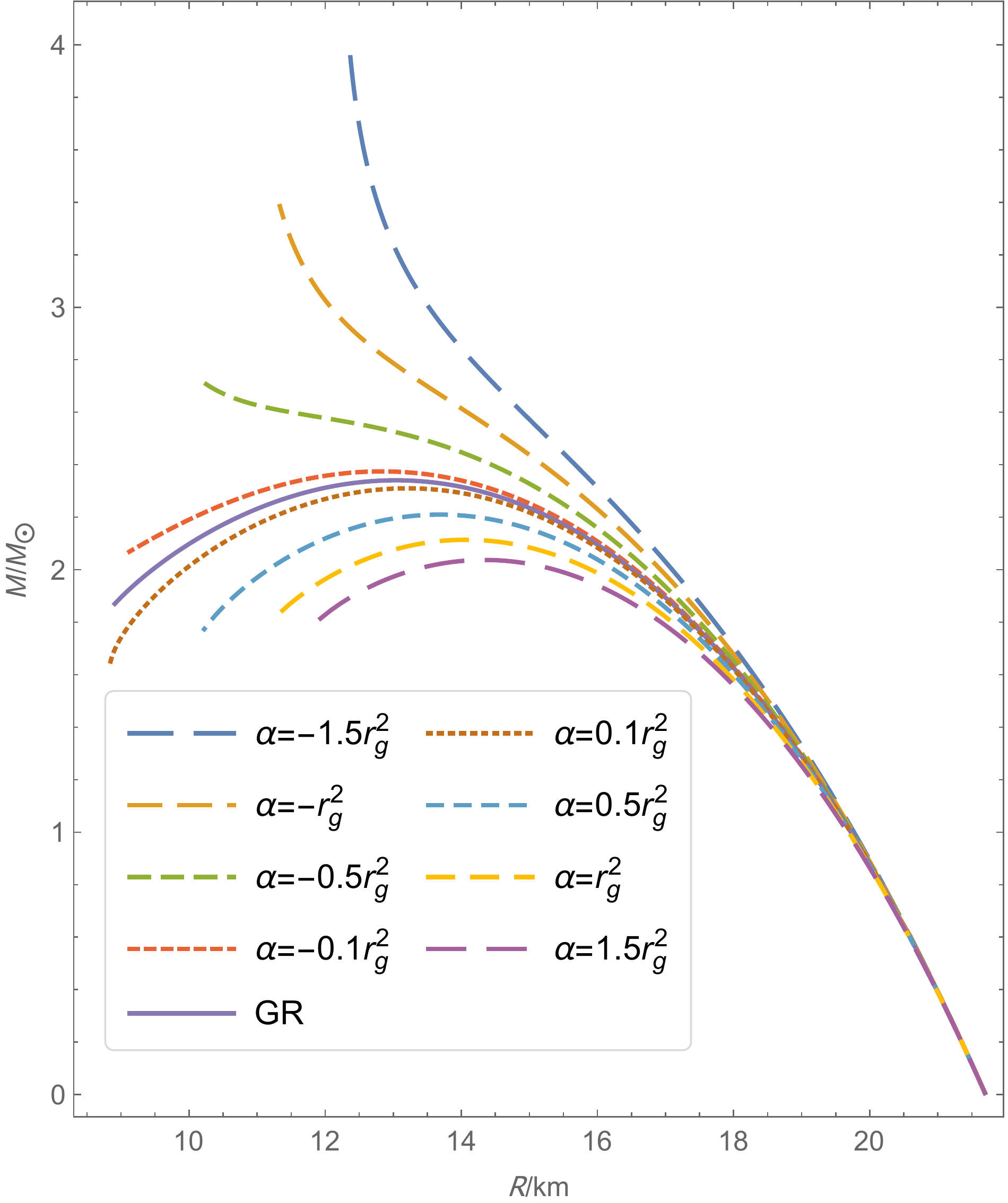}
    \caption{Mass-to-radius curves for polytropic fluids with $\gamma=2$ and $f(Q)=Q+\alpha Q^2$ with different values of $\alpha$.}
    \label{fig:MR}
\end{figure}

\begin{figure}[htpb]
    \centering
    \includegraphics[width=0.5\linewidth]{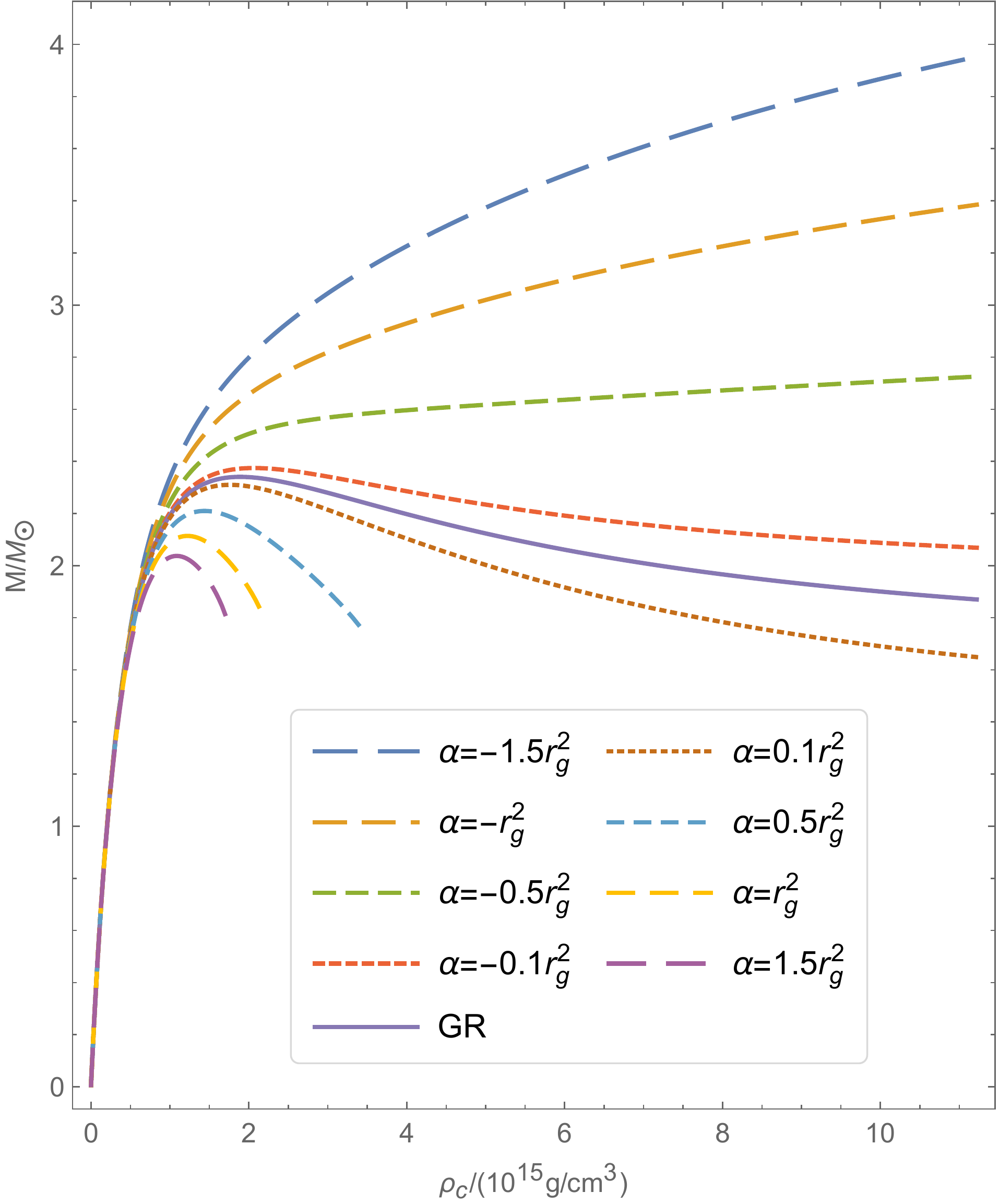}
    \caption{Relations between the star masses and central densities with $\gamma=2$ and $f(Q)=Q+\alpha Q^2$ with different values of $\alpha$.}
    \label{fig:Mrho}
\end{figure}

Numerical procedure shows that the system given by Eqs. \eqref{eomsph} and \eqref{dp} is able to converge to solutions of compact stars.
Figure \ref{fig:profile} illustrates the profiles of the energy density $\rho$,
nonmetricity scalar $Q$, and the metric $\xi,\zeta$,
where $r_g=G M_\odot/c^2\simeq 1.48\times10^5\:\text{cm}$,
and we have set the central density $\rho(0)=\rho_c$ to be the typical density of neutron matter $\rho_c=10^{15}\text{g/cm}^3$ \cite{prakash}.
One can see that the distribution of matter is indeed confined in a finite radius,
and the nonmetricity scalar $Q$ reaches zero at this radius and matches the external solution.
The metric functions $\xi$ and $\zeta$ are also connected regularly at this radius.
In Fig. \ref{fig:MR}, we illustrate the mass-radius relation for various model parameter $\alpha$.
The corresponding mass-$\rho_c$ curves are shown in Fig. \ref{fig:Mrho}.
The stellar radii generally decrease as the central densities increase.
For lower central density, the radii of the stars with various $\alpha$ approach the same value,
which coincides with the Newtonian stellar radius given by the Lane-Emden equation for polytropic fluids with $\gamma=2$,
i.e.,
\begin{equation}
    \label{LEradius}
    \mathcal R_\text N=\pi\sqrt{\frac{k\gamma c^2}{4\pi G(\gamma-1)}}\simeq 21.703\:\text{km}.
\end{equation}
For higher central density, the numerical procedure continues till it cannot converge stably.
Thus, beyond the curves, either the system described by Eqs. \eqref{eomsph} and \eqref{dp} is unstable or a more realistic equation of state is required.

For positive $\alpha$, as well as negative $\alpha$ but with relatively small $|\alpha|$ (see the $\alpha=-0.1 r_g^2$ curve),
the curves display similar behaviors to the GR case.
A critical configuration can be found such that the stellar mass reaches its maximum.
If the ratio between pressure and energy density goes beyond this critical point, the structure of the star may become unstable.
For negative $\alpha$ with sufficiently large $|\alpha|$,
such critical point cannot be found.
An analogous pattern is reported in $f(T)$ gravity \cite{PhysRevD.98.064047}.
Figures \ref{fig:MR} and \ref{fig:Mrho} suggest that for $\alpha>0$ less stellar mass than GR is allowed,
while for $\alpha<0$ much more mass can be supported.

\section{Conclusion and discussions}
\label{conclusion}
Although the symmetric teleparallel formulation considers a different connection than the Levi-Civita one,
the manifold itself remains the same from a mathematical point of view.
Therefore, the different bases of geometric descriptions of gravity, i.e. GR, TEGR and STGR, are in fact equivalent to each other.
The vanishing of curvature and torsion forces the connection $\Gamma$ of $Q$-based theories to be purely inertial.
That is, $\Gamma$ should not include any information of gravity and
hence can be determined from the corresponding spacetime where gravity is canceled.
On this basis, the covariant $f(Q)$ gravity can be constructed
and the Lorentz symmetry of the theory is respected in this formalism.
These constitute the theoretical basis of our exploration on the application of $f(Q)$ gravity to the spherically symmetric scenarios.

As heuristic examples we have investigated the external and internal solutions of spherically symmetric objects in $f(Q)$ gravity.
For the exterior spacetimes with and without electromagnetic field,
we find that the $f(Q)$ gravity always results in the same external solutions as in GR,
i.e., the RNdS and SdS solutions, respectively.
The different forms of $f(Q)$ may, at background level, only affect the constant term acting as the cosmological constant.
This coincidence comes possibly from the high symmetry of the spherical configurations.
It can be checked that in the rotating case, the nonmetricity scalar $Q$ does not vanish for the Kerr metric.
Therefore, it is expected that more varieties should appear in the axially symmetric configurations of $f(Q)$ gravity.
Nonetheless, the coincidence of the external solutions with GR may ensure
that $f(Q)$ gravity does not present any extra forces locally and is able to pass the weak field tests as GR,
while in cosmology, it can give the explanation to the accelerated expansion of the Universe
\cite{PhysRevD.100.104027,Lu:2019hra,PhysRevD.101.103507,PhysRevD.102.024057,PhysRevD.102.124029,Barros:2020bgg}.
At perturbation level, the external solutions no longer coincide due to
the effects of $f(Q)$ that emerge in the perturbed equation of first order.

The well-defined external solutions in spherically symmetric configurations also provide
reliable junction conditions and hence boundary conditions to solve the internal field equations of the compact star.
Within the star, models with different forms of $f(Q)$ do not coincide with each other and do have their effects on the interior structures.
We demonstrate these effects of $f(Q)$ by considering a quadratic model of $f(Q)=Q+\alpha Q^2$
and a maximally stiff model of polytropic fluid with a first adiabatic coefficient $\gamma=2$ \cite{PhysRevD.98.064047}.
Through numerical procedures we find that models with a negative $\alpha$ may support more matter for the compact star than in GR,
while with a positive $\alpha$ less matter can be contained.
This may be interpreted in the following way.
In the interior region of a compact star,
$f(Q)$ model with a positive modification exerts a stronger gravity than STGR for a given amount of matter.
Thus, at the same pressure level, e.g., degeneracy pressure of nuclear particles, less matter can be supported in such a case.
On the other hand, a negative modification may act as diminishment of the gravity,
or, if considered as a geometric fluid, provide a pressure to resist the gravity;
hence, more matter can be supported.
For a negative $\alpha$ with sufficiently large $|\alpha|$,
no critical configuration of maximum mass could be found,
which may suggest possible instability.
Yet a full-fledged analysis that encompasses temporal evolution of the system should be considered.

Another issue may come from the EOS.
For simplicity, we have adopted a polytropic EOS.
More realistic EOS's are required if one wishes to compare predictions of compact stars, e.g., neutron stars
in $f(Q)$ gravity with the astrophysical observations.
A recent report of GW190814 \cite{Abbott:2020khf} indicates that a compact object, possibly a neutron star,
of $2.6M_\odot$ may exist.
Then instead of an EOS posing challenges for the nuclear physics,
modifications in the theory of gravitation such as $f(Q)$ gravity may be able to offer an alternative explanation to it.
\section*{Acknowledgement}
\label{ackn}
This work is supported by the National Science Foundation of China under Grants No. 11847080.

\bibliography{ref}
\end{document}